# Non-photorealistic image rendering with a labyrinthine tiling

A. Sparavigna[1] and B. Montrucchio[2]
[1]*Physics Department, Politecnico di Torino*
[2]*Control and Computer Engineering Department, Politecnico di Torino*
*Corso Duca degli Abruzzi 24*
*Torino, I-10129, Italy*

The paper describes a new image processing for a non-photorealistic rendering. The algorithm is based on a random generation of gray tones and competing statistical requirements. The gray tone value of each pixel in the starting image is replaced selecting among randomly generated tone values, according to the statistics of nearest-neighbor and next-nearest-neighbor pixels. Two competing conditions for replacing the tone values - one position on the local mean value the other on the local variance - produce a peculiar pattern on the image. This pattern has a labyrinthine tiling aspect. For certain subjects, the pattern enhances the look of the image.

## 1. INTRODUCTION

Photorealistic processing of images is not always the best vehicle for communicate information, because the success of the applied processing depends on the intention behind the communication. This processing is often required for scientific and forensic imagery, where photorealistic processing algorithms can remove noise or enhance object outlines in the image scene. However, it is not obvious that a photorealistic processing is always to be preferred. Let us consider for instance the hand-drawing, clearly a non-photographic imagery. The idea that these illustrations can be better to explain a scene than photographic plates is quite natural. Hand-drawing helps to illustrate complex phenomena, omitting details and proposing only fundamental objects, sometimes with a symbolic representation.

The goal of the non-photorealistic rendering (NPR) (for a recent literature survey[1-4]) is the development of algorithms for generating or processing images that embody the following qualities: emphasis of selected features, suppression of unimportant details, use of stylization to suggest emotional structures. Some NPR algorithms are devoted to produce pseudo hand-drawing images[5]. Other techniques are giving interesting aesthetic results[6] working with algorithms for visualizing the vector fields. The vector field visualization, used also for scientific purposes[7,8], is mainly based on line integral convolutions (LICs). LIC and related techniques use stylization to suggest the structure of the vector field, as an hand-drawing stylization of the field can do.

We show here an image processing based on a statistical approach to obtain a non-photorealistic rendering, different from vector field visualization, or from other techniques such as mosaic or tessellation textures. It is based on the use of statistical parameters such as mean value and variance and gives on the processed image a labyrinth tiled texture, which is not governed by casualty but able to follow the outlines of the objects in the image scene. The stylization we propose is then interesting non only for aesthetic purposed but for the enhancement of visibility of certain features of images.

## 2. LABYRINTH-LIKE TILING

The rendering we propose is a random pixel tone generation performed with two competing statistical requirements. This is giving a peculiar pattern on the image, producing a tiling of the image with a labyrinth-like texture. As in many physical and chemistry phenomena, where competing conditions are acting on the system, complex patterns are displayed[9-11]. It is quite usual to encounter stripes, fingerprints or labyrinths in the optical investigations of solid surfaces or of thin liquid crystal films. Competing conditions give two or more possible local orientations or configurations of the material, revealed by complex patterns in microscopy observations.

Before discussing the algorithm, let us show the result of the image processing, to illustrate the pattern formed by the algorithm, on a photographic image. In the Figure 1, the source image and its rendering are compared. The texture has a labyrinth structure, which is not arbitrary but is following the contours of the objects, in this case of eyes, eyebrows, chin, etc. In Figure 2 another example of rendering on the portrait of Benjamin Franklin.

The algorithm is based on a statistics manipulations of the image. The statistical approach means that decisions and choices of the algorithm are based on statistical parameters such as mean values, variances and higher moments[12-14]. Before discussing the algorithm, let us observe that an image processing algorithm based simply on the global mean value and variance of the image, that is on the mean value and variance evaluated on all the pixels forming the image, can be misleading. Each part of the scene possesses a local mean value and variance. These local values can be rather different.

For each pixel in the image then, we use a local mean value and variance defined in the following way:

$$M(x,y) = \frac{1}{N} \sum_{\substack{n.n. \\ n.n.n.}} g(x,y) ;$$

$$V(x,y) = \frac{1}{N} \sum_{\substack{n.n. \\ n.n.n.}} [g(x,y) - M(x,y)]^2 \qquad (1)$$

where (x,y) are the co-ordinates of the pixel under consideration in the image frame. Function g(x,y) is the gray tone of the pixel and then a function of the pixel position. Additions are done on the nearest neighbor (n.n.) and next nearest neighbor (n.n.n.) pixels of the place (x,y). N is the number of nearest neighbor and next nearest neighbor pixels. Starting for the left upper corner of the image we check all the pixels of the image. We can choose to keep the gray tone value g(x,y) of the pixel (x,y) under test if :

$$\Re_1 = \frac{|g(x,y) - M(x,y)|}{M(x,y)} \leq t \qquad (2)$$

where $t$ is a fixed threshold value. Eq.2 is comparing the pixel gray tone with the local mean value of the neighbouring pixels. For our image rendering, we choose a value threshold value of 12%.

If the ratio $\Re_1$ exceeds the fixed threshold, the value of the gray tone is changed with a new value evaluated in the following way. The values of the gray tones range from 0 to 255: let us subdivide this range in three (for example) approximately equally spaced ranges, with lower and upper values, $L_{inf}^i$ and $L_{sup}^i$, with index $i$ ranging from 1 to 3. For neighbor intervals: $L_{sup}^i = L_{inf}^{i+1}$. For the first interval, it is better to take as $L_{inf}^1$ a low value different from zero. The gray tone g(x,y) of the pixel is belonging to one of these ranges. The new gray tone g*(x,y) of the pixel is given according to the range by means of the following expression:

$$g*(x,y) = T_{inf}^i + R\left(T_{sup}^i - T_{inf}^i\right) \qquad (3)$$

where R is a random number going from 0 to 1. $T_{inf}^i, T_{sup}^i$ are gray tones which can be coincident with the values $L_{inf}^i, L_{sup}^i$ or different to enhance the rendering of the image. Position in Eq.2 is performed again and if Eq. 2 is true, the new value g*(x,y) substitutes the old value g(x,y) of the gray tone. If it is not so, another random number is generated and the procedure continues till condition (2) is satisfied or if the number of iterations exceeds a fixed value. Let us note that since the value of gray tone of the pixel is changed, the local environment of the pixel is also changed: the mean value M(x,y) and variance V(x,y) of Eq.(1) acquire new values as the processing evolves in checking all pixels of the image.

To obtain the tiling pattern in Fig.1 and 2, a competing condition must be inserted in the algorithm. In fact, the pixel tone substitution that we have previously described gives a local smoothing of the image, removing random noise. The substitution $g(x,y) \to g*(x,y)$ alone is then not able to create the labyrinthine pattern. We must insert a condition competing against this smoothing, a condition then able to strongly enhance the local variance of pixel tones. This is accomplished with the following control.

The substitution $g(x,y) \to g*(x,y)$ is not applied if we find the following condition:

$$\Re_2 = \frac{\sqrt{V(x,y)/N}}{M(x,y)} > V \qquad (4)$$

verified at the pixel (x,y). A threshold $V$ of 50% is used for patterns in Fig.1 and 2. A high value of the ratio $\Re_2$ means that the pixel (x,y) is belonging to a region where the gray tone function has a strong local gradient.

The role of the value of the threshold $V$ in giving the tiling is illustrated in Figure 3, where the results on the pattern for a slight change in the values of $V$ are shown: on the left the threshold is fixed at 47% and on the right at 53%. In the middle, the threshold value is of 50%. The figure display that a different value of the threshold parameter gives a different tiling on the image. Trials on several images demonstrate that there is only a small range of values for the threshold $V$ where the labyrinth pattern is displayed on the processed image.

3. ENHANCING IMAGES.

The ratio $\Re_2$ in condition (4) depends on the local gradient of the gray tone function g(,y). When the pixel is very near or is on the contour of an object in the image, the ratio $\Re_2$ increases. To verify that position (4) is acting in enhancing the edges, the algorithm was applied several times to the image. Fig.4 is a sequence of images obtained with this iteration procedure. The last image of the sequence (Fig. 4 (d)) looks like an image resulting from the application of an edge detection filter.

The stylization here proposed is able to give in certain circumstances, a better appreciation of a scene. An example of the algorithm applied to a landscape (Fig.5), can be useful to stress this possibility. In Fig.5, the details of the small branches of the tree are suppressed but the shape of the main branches is strongly enhanced by the labyrinth tiling. In this case then, the morphology of the tree is put in evidence by the rendering: it seems that the growth of the branches in the image is following the labyrinthine pattern. The intrinsic geometric structure (the tree) in the image is favored by the stylization of the rendering algorithm.[15] This is a consequence of position (4): as in the sequence of Fig.4, we have a slight enhance of the edges.

As we have argued at the beginning, one of the goals in non realistic rendering is to stimulate sensations, with a certain enhancing of patterns in the image scene. If the rendering algorithm turns out to be good for this purpose is, to a certain extent, a rather subjective conclusion. We are now working to evaluate in a quantitative manner the features of the labyrinthine tiling.

REFERENCES

[1] F. Durand, *An Invitation to Discuss Computer Depiction*, Proc. Int. Symp. On Non-photorealistic Animation and Rendering, NPAR02, ACM Press, pp. 111-124 (2002).
[2] M. Tory and T Moller, IEEE Trans. Visualization and Computer Graphics 10, 72-84 (2004).
[3] T. Strothotte and S. Schlechtweg, *Non-Photorealistic Computer Graphics: Modeling, Rendering, and Animation*, Morgan Kaufman Publisher, San Francisco (2002).
[4] C. Reynolds, *Stylized Depiction in Computer Graphics*, an Internet report (2006) http://www.red3d.com/cwr/npr/, 2006.
[5] I. Buck, A. Finkelstein, C. Jacobs, A. Klein, D. H. Salesin, J. Seims, R. Szeliski, and K. Toyama, *Performance-driven hand-drawn animation*, Proc. Int. Symposium on Non-photorealistic animation and rendering, NPAR00, ACM Press, pp.101-108 (2000).
[6] C-M. Wang and J-S. Lee, J. Information Science and Engineering 20, 923-948 (2004).
[7] A.Sparavigna, A.Sanna, B.Montrucchio and A.Strigazzi, Liq. Cryst. 26, 1467-1478 (1999).
[8] B. Montrucchio, P. Montuschi, A. Sanna, and A. Sparavigna, Computers & Graphics 25, 847-855 (2001).
[9] J.P. Gollub and J.S. Langer, Rev. Mod. Phys. 71, s396-s403 (1999).
[10] C. Bowman and A. C. Newell, Rev. Mod. Phys. 70, 289-301 (1998).
[11] A. J. Koch and H. Meinhardt, Rev. Mod. Phys. 66, 1481-1507 (1994).
[12] B.Montrucchio, A.Sparavigna and A.Strigazzi, Liq. Cryst. 24, 841-852 (1998).
[13] B.Montrucchio, A.Sparavigna, S.I.Torgova and A.Strigazzi, Liq. Cryst. 25, 613-620 (1998).
[14] A.Sanna, B.Montrucchio and A.Sparavigna, Pattern Rec. Lett. 20, 183-190 (1999).
[15] The reader can find images with a high resolutions at the web page staff.polito.it/ amelia.sparavigna/rendering.

FIGURE CAPTIONS

FIG.1: Example of labyrinthine tiling on a photographic image.

FIG.2: Labyrinthine tiling on the portrait of Benjamin Franklin.

FIG.3: The role of the threshold value on the pattern. From left to right the threshold value is 47%, 50% and 53%.

FIG.4: An iterative sequence of the algorithm. The last image (d) looks like the image that can be obtained with an edge detection filter.

FIG.5: A example of the rendering of a landscape. Details of the small branches of the tree are suppressed but the shape of the main branches is strongly enhanced by the labyrinth tiling (source image: Norman Walsh 2006). To appreciate the rendering see the figure with a better resolution at staff.polito.it/ amelia.sparavigna/rendering.

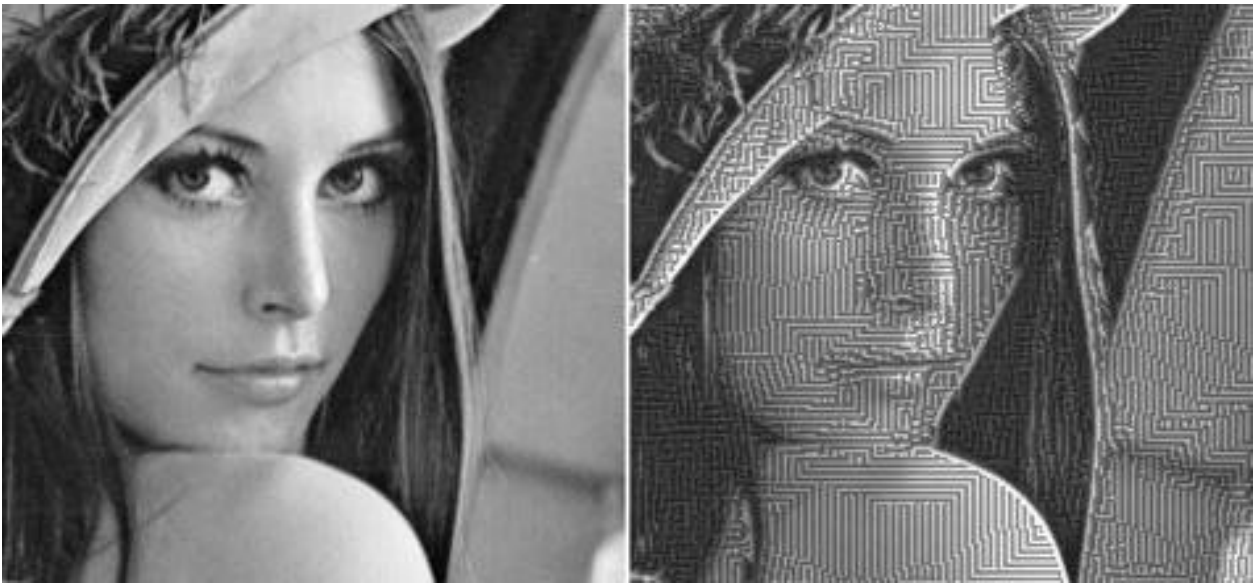

FIGURE 1

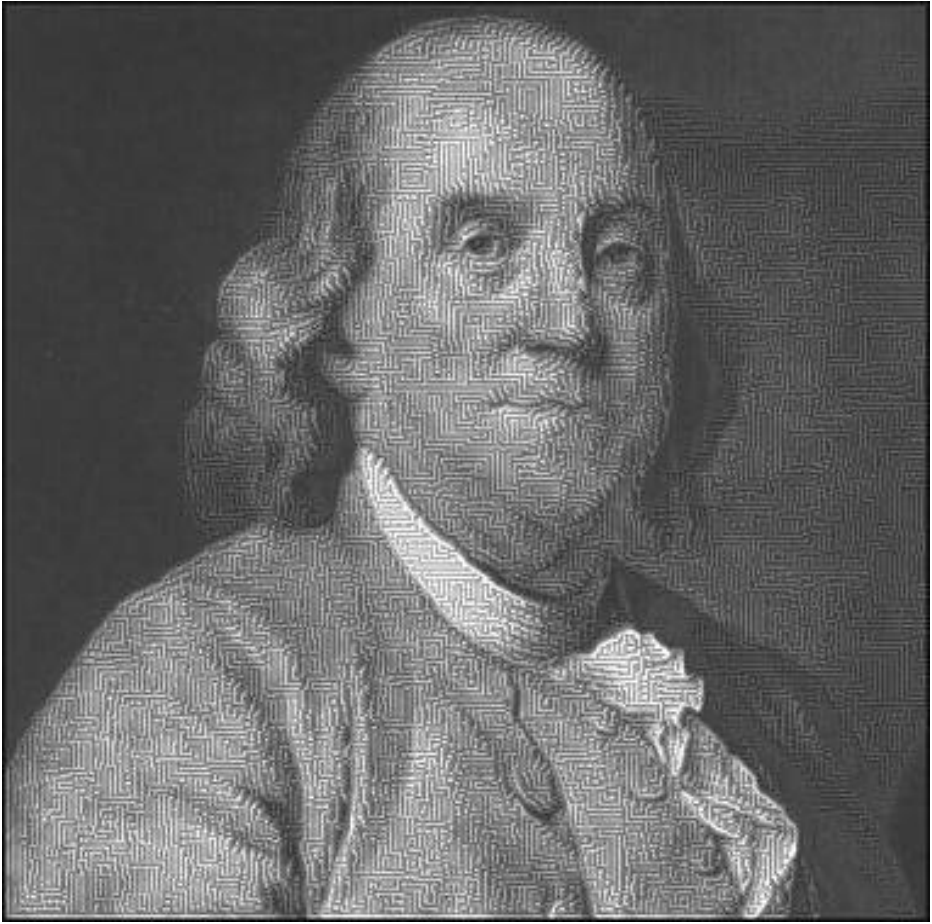

FIGURE 2

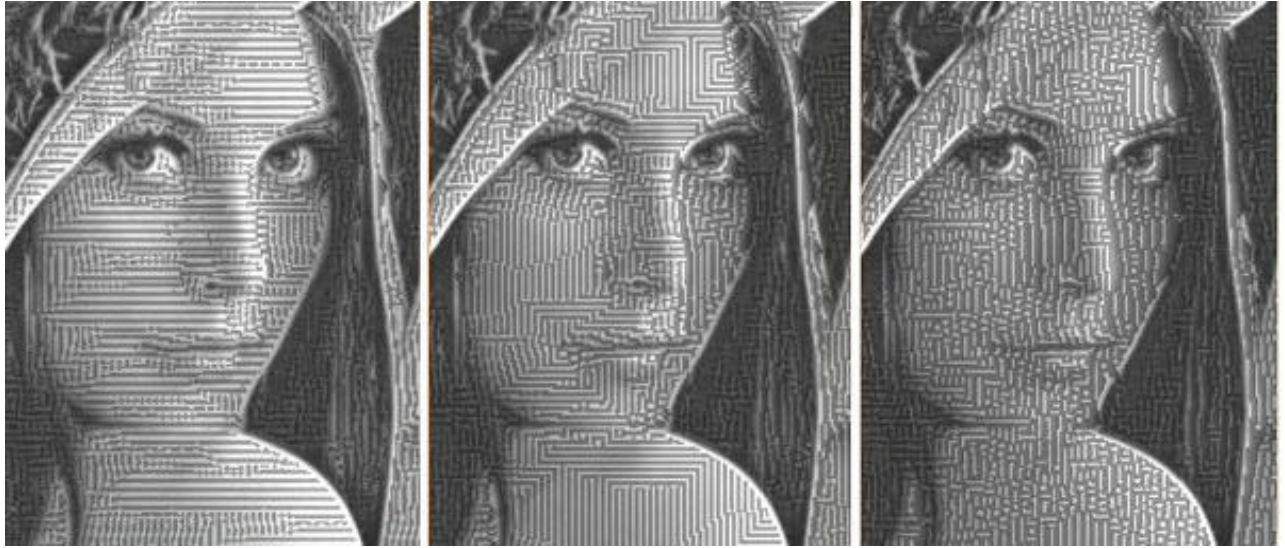

FIGURE 3

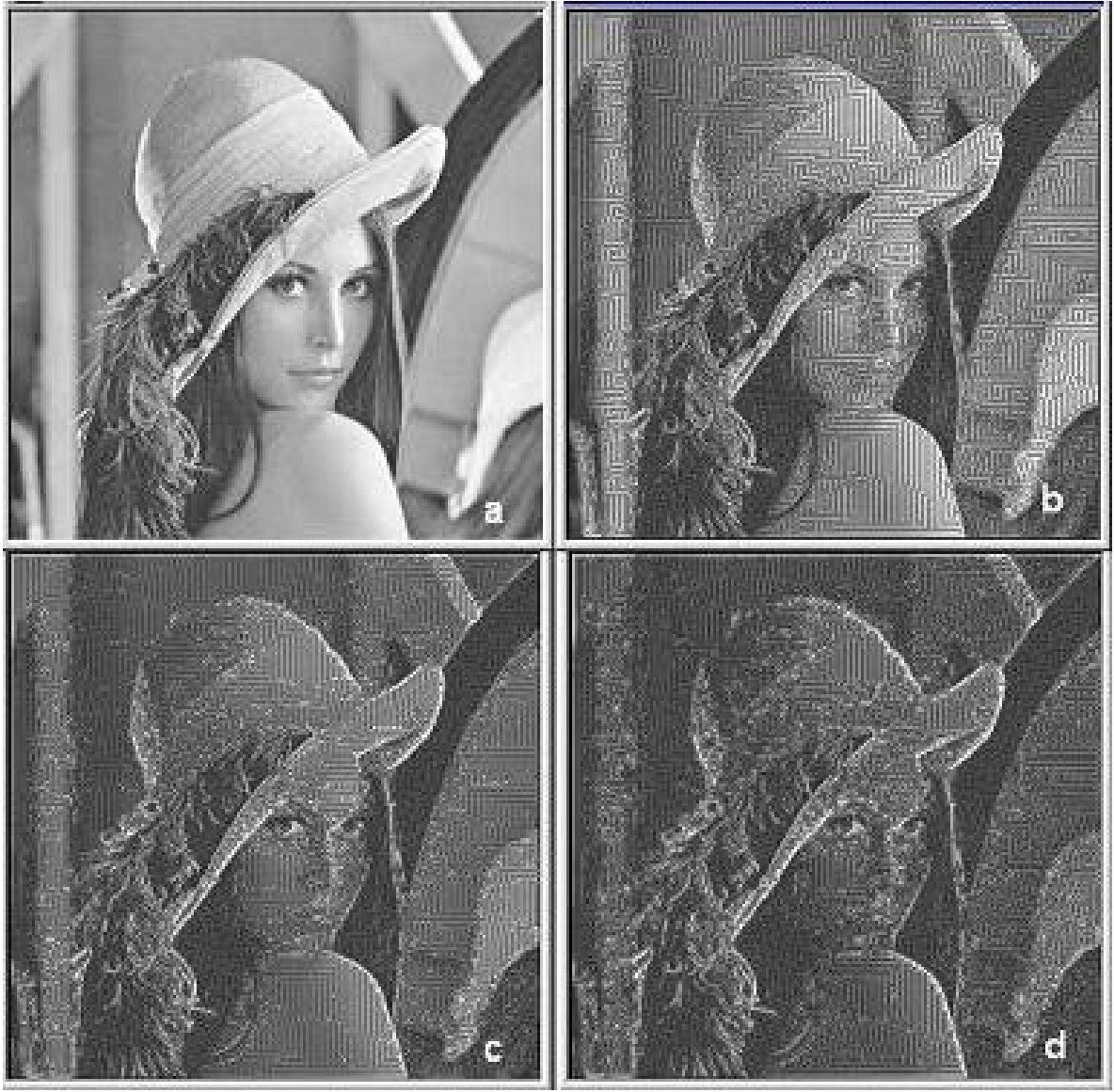

FIGURE 4

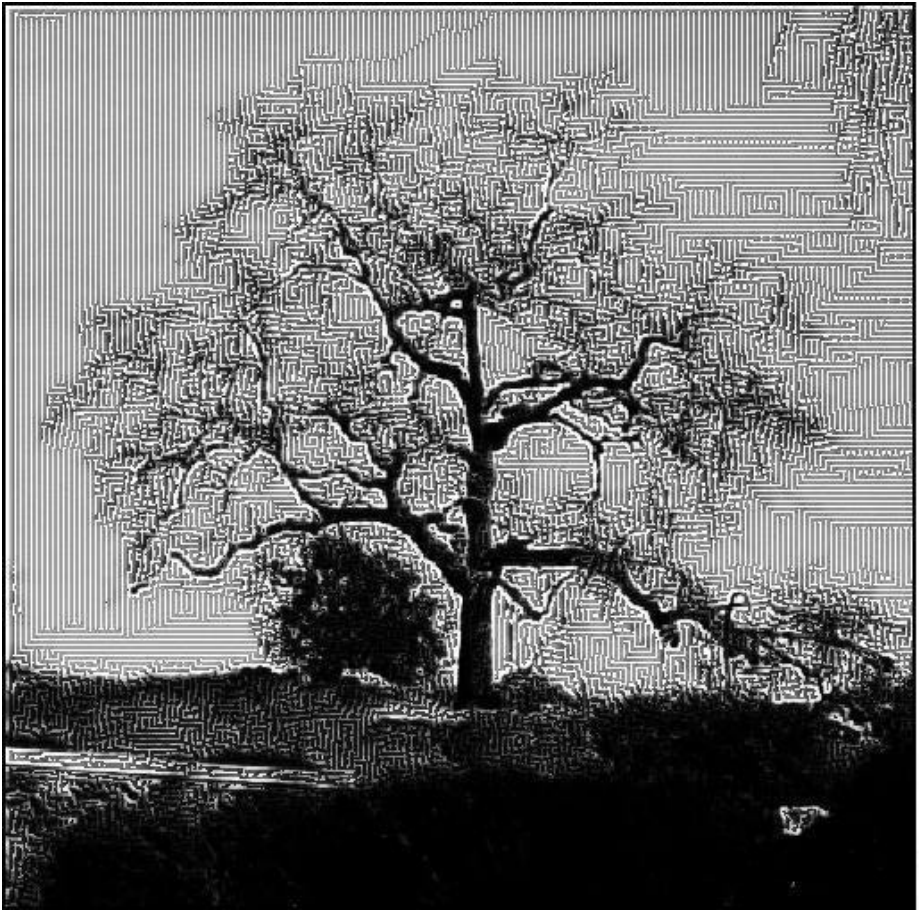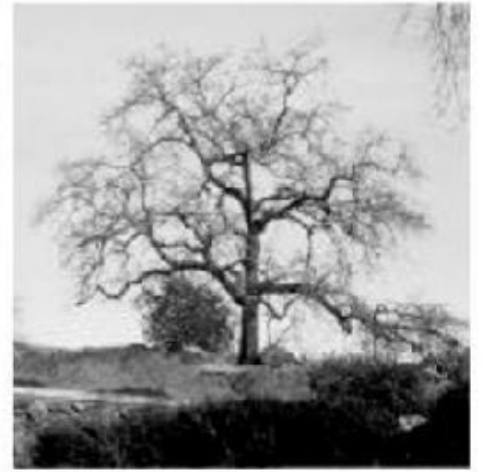

FIGURE 5